\documentstyle[12pt,world_sci,epsf]{article}

 1
\font\elevenrm=cmr10 scaled\magstep 1
\font\elevenit=cmti10 scaled\magstep 1

\font\twelvebf=cmbx12

\renewenvironment{thebibliography}[1]
 { \elevenrm
   \begin{list}{\arabic{enumi}.}
    {\usecounter{enumi} \setlength{\parsep}{0pt}
     \setlength{\itemsep}{3pt} \settowidth{\labelwidth}{#1.}
     \sloppy
    }}{\end{list}}

\newcommand{\vub}{V_{ub}}
\newcommand{\vcb}{V_{cb}}
\newcommand{\btou}{b\to u\ell\nu}
\newcommand{\btoc}{b\to c\ell\nu}
\newcommand{\upsfs}{\Upsilon(4S)}
\newcommand{\gevc}{GeV$/c$}
\newcommand{\mevc}{MeV$/c$}

\newcommand{\pilv}{\pi\ell\nu}
\newcommand{\rholv}{\rho\ell\nu}
\newcommand{\omeglv}{\omega\ell\nu}

\newcommand{\rhomlv}{\rho^-\ell^+\nu}
\newcommand{\rhonlv}{\rho^0\ell^+\nu}
\newcommand{\omelv}{\omega\ell^+\nu}
\newcommand{\pimlv}{\pi^-\ell^+\nu}
\newcommand{\pinlv}{\pi^0\ell^+\nu}

\newcommand{\rhomlvc}{\rho^\mp\ell^\pm\nu}
\newcommand{\rhonlvc}{\rho^0\ell^\pm\nu}
\newcommand{\omelvc}{\omega\ell^\pm\nu}
\newcommand{\pinlvc}{\pi^0\ell^\pm\nu}
\newcommand{\pimlvc}{\pi^\mp\ell^\pm\nu}

\newcommand{\bbpi}{{\cal B}(B^0\to\pimlv)}
\newcommand{\bbrho}{{\cal B}(B^0\to\rhomlv)}
\newcommand{\gampi}{\Gamma(B^0\to\pimlv)}
\newcommand{\gamrho}{\Gamma(B^0\to\rhomlv)}

\newcommand{\emiss}{E_{miss}}
\newcommand{\pmiss}{\vec{p}_{miss}}
\newcommand{\dele}{\Delta E}
\newcommand{\mb}{m_B}
\newcommand{\pnu}{\vec{p}_\nu}

\newcommand{\pel}{\vec{p}_\ell}
\newcommand{\mm}{M_{miss}^2}
\newcommand{\npi}{N_{\pi^\pm\ell^\mp\nu}}
\newcommand{\nrho}{N_{\rho^\pm\ell^\mp\nu}}
\newcommand{\e}[1]{\times10^{#1}}
\newcommand{\thepil}{\theta^*_{\pi\ell}}
\newcommand{\cospil}{\cos\thepil}
\newcommand{\invfb}{fb$^{-1}$}
\newcommand{\bbar}{B\bar{B}}

\newcommand{\etal}{{\it et al.}}
\newcommand{\plb}[1]{Phys. Lett. {\bf B#1}}
\newcommand{\prl}[1]{Phys. Rev. Lett. {\bf #1}}
\newcommand{\npb}[1]{Nucl. Phys. {\bf B#1}}
\newcommand{\prd}[1]{Phys. Rev. {\bf D#1}}
\newcommand{\zpc}[1]{Z. Phys. {\bf C#1}}
\newcommand{\nim}[1]{Nucl. Instrum. Methods Phys. Res. Sect. A {\bf #1}}
\newcommand{\beq}{\begin{equation}}
\newcommand{\eeq}{\end{equation}}

\newcommand{\eo}{\hspace*{0.5em}}
\newcommand{\et}{\hspace*{0.5em}}

\begin{document}
%%% the title page
\begin{titlepage}
\begin{center}
\vspace*{1.0 truecm}
{\elevenrm \hfill UR-1427}\\
{\elevenrm \hfill June, 1995}\\
\vspace*{2.5 truecm}
{\twelvebf PRELIMINARY RESULTS FOR EXCLUSIVE \boldmath$\btou$\\
DECAYS FROM CLEO}\\
\vspace{1 truecm}
{\elevenrm Lawrence Gibbons}\\
{\elevenit University of Rochester, Rochester, NY  14627}\\

\newlength{\abskip}
\setlength{\abskip}{4.0 truecm}
\addtolength{\abskip}{-6\baselineskip}
\vspace*{\abskip}

\begin{abstract}
{ \noindent
A preliminary analysis of exclusive $\btou$ decays to the final states
$\pimlvc$, $\pinlvc$, $\rhomlvc$, $\rhonlvc$ and $\omelvc$ based on $2.2\e{6}$
$\bbar$ decays collected at CLEO is presented.  We have measured the first
exclusive $\btou$ branching fraction $\bbpi=[1.19\pm0.41\pm0.21\pm0.19]\e{-4}$
($[1.70\pm0.51\pm0.31\pm0.27]\e{-4}$), with the ISGW (WSB) model used for
efficiency determination.  A 90\% C.L. upper limit on $\bbrho$ similar to the
previous CLEO limit is obtained.  The ratio $\gamrho/\gampi<3.4$ at the 90\%
confidence level for both the ISGW and WSB models.  This ratio provides some
discrimination between form factor models.
}
\end{abstract}
\vspace*{2.0 truecm}
{\elevenit To appear in the Proceedings of the XXXth Rencontres de Moriond\\
``Electroweak Interactions and Unified Theories'',\\
Les Arcs, France, March, 1995}
\end{center}
\end{titlepage}

%%% Now, go to 1 1/2 line spacing
%%\renewcommand{\baselinestretch}{1.4} \small  \normalsize

% start of text

\section{Introduction}
This talk will focus on a preliminary CLEO analysis of $\btou$ decays to the
exclusive final states $\pilv$, $\rholv$ and $\omeglv$.  The ultimate goal of
this analysis is to improve our knowledge of $|\vub|$.
ARGUS \cite{bb:argus_inclusive} and CLEO \cite{bb:cleo_inclusive} have already
demonstrated that $|\vub|>0$ by examining the inclusive lepton momentum
spectrum from $B$ decays at the $\upsfs$. They observe events beyond 2.4 \gevc,
which is kinematically forbidden for the copious $\btoc$ processes, but is
still accessible to $\btou$ decays.  While these analyses clearly establish an
excess in this endpoint region, and hence that $|\vub|>0$, extracting a
reliable value of $|\vub|$ is difficult because of the theoretical uncertainty
in extrapolating from the observed rate in the endpoint region to the total
$\btou$ rate.  Values of $|\vub/\vcb|$ obtained from these analyses are now in
the 7\% to 11\% range, with the theoretical uncertainty dominating.

\section{Exclusive $\btou$}
An alternate route to $|\vub|$ is through the study of exclusive $\btou$
channels. The best previous information concerning such channels is the
upper limit set by CLEO \cite{bb:UCSBlimit} in the combined modes $\rhomlv$,
$\rhonlv$ and $\omelv$. The CLEO result
corresponds to an upper limit of $\bbrho<3.2\e{-4}$ at the 90\% confidence
level (ISGW model \cite{bb:isgw}).

The preliminary analysis presented here studies the two pseudoscalar modes
$\pimlv$ and $\pinlv$, the three vector modes $\rhomlv$, $\rhonlv$ and
$\omelv$, and the charge conjugate modes.  At a fixed $|\vub|$, the existing
form factor models predict a wide range of partial widths for these modes, as
Table~\ref{tab:exclpred} shows.  Unfortunately, measured branching
fractions depend on the form factor model used to evaluate the experimental
efficiencies, as does the extraction of $|\vub|$.  We therefore need to
discriminate between the different models.

\begin{table}[b]
\centering
\caption{Predictions for the exclusive partial widths $\gampi$ and $\gamrho$
and the ratio $\gamrho/\gampi$.  The partial width units are
$10^{12}|\vub|^2$ sec$^{-1}$.}
\label{tab:exclpred}
\begin{tabular}{lccc} \hline\hline
Model                    & $\gampi$     & $\gamrho$ & $\gamrho/\gampi$\\ \hline
WSB \cite{bb:WSB}        & 6.3 -- 10.0 &  18.7 -- 42.5 & 3.0 -- 4.3 \\
KS \cite{bb:KS}          & 7.25        &  33.0         & 4.6 \\
ISGW \cite{bb:isgw}      & 2.1         &  8.3          & 4.0 \\
ISGW II \cite{bb:ISGWII} & 9.6         & 14.2          & 1.5 \\ \hline
\end{tabular}
\end{table}

The ratio $\gamrho/\gampi$ provides one means of discrimination.  Because the
$\pilv$ rate is helicity-suppressed when the daughter meson is at rest in the B
meson rest frame (at $q^2_{max}$), where the form factors for the decay are
largest, while the $\rholv$ rate is not, we expect the ratio to be larger than
one. The exact value for the ratio will depend on the $q^2$-behavior of the
form factors.  In Table~\ref{tab:exclpred}, we see that the predictions of the
ratio span a fairly broad range, so the ratio should prove useful.

\section{Neutrino ``Measurement'' and Exclusive $\btou$}
Experimentally, semileptonic decays are troublesome because of the undetected
neutrino.  This analysis takes advantage of the excellent hermeticity and
resolution of the CLEO II detector located at the Cornell Electron Storage Ring
(CESR) to obtain information about the neutrino in semileptonic $\btou$
decays.  Three concentric tracking devices provide a momentum resolution of
$\sigma_p/p=0.005\oplus 0.0015p$ ($p$ in \gevc), while covering 95\% of the
$4\pi$ solid angle. The CsI calorimeter located inside of the CLEO solenoid
provides an energy resolution well approximated by
$\sigma_E/E=0.019+0.0035/E^{0.75}-0.001E$ ($E$ in GeV), while covering 98\% of
$4\pi$.  The detector is described in detail elsewhere \cite{bb:nim}. This
analysis is based on a data sample with a luminosity of $2.09$ \invfb\
(about $2.2\e{6}$ $\bbar$ decays).

The underlying idea is very simple: the $B\bar{B}$ system is at rest at CLEO
and the beam energy is known very precisely, so we can ``measure'' the neutrino
four momentum by measuring the missing energy and momentum of an event.  We
define
\begin{eqnarray}
\emiss & \equiv & 2E_{beam} - \sum_i E_i \\
\pmiss & \equiv & -\sum_i \vec{p}_i,
\label{eq:missdef}
\end{eqnarray}
where the index $i$ runs over all charged tracks and all showers in the
calorimeter that pass cuts designed to reject false tracks and spurious
showers from hadronic interactions.

In events with no extra missing particles,
$\pmiss$ can be reliably associated with the momentum $\pnu$ of the signal
mode neutrino. The $\btou$ decay can then be fully reconstructed:  the energy
difference $\dele\equiv E_{beam} - (E_h+E_\ell+|\pnu|)$, where $h$ is the
candidate hadron, should be zero, and the beam-energy constrained mass
$\mb\equiv\sqrt{E_{beam}^2-|\vec{p}_h+\pel+\pnu|^2}$ should reconstruct at the
$B$ mass.  Signal events that are reconstructible show resolutions of
approximately 260 MeV on $\emiss$ and 110 MeV on $|\pmiss|$.

Signal events with particles missing in addition to the neutrino usually fail
the reconstruction criteria. On the other hand, those background events that
pass the criteria do so because they have extra particles missing.
Consequently, we reject events with multiple leptons or a non-zero total charge
because they indicate a second neutrino or a missed charged particle,
respectively.  Most remaining events with extra missing particles are
eliminated by requiring that $\mm\equiv\emiss^2 - |\pmiss|^2$ be consistent
with zero. The criterion $\mm/2\emiss<350$ MeV is used since the $\mm$
resolution varies approximately as $2\emiss\sigma_{\emiss}$.

Continuum background is suppressed using standard event shape variables.  The
$\btou$ processes are enhanced over $b\to c$ by requiring the leptons to have
momenta larger than 1.5 \gevc\ (2.0 \gevc) in the $\pilv$
(vector) modes.  The lower cut is used in the $\pi$ modes because
these modes are expected to have a softer lepton momentum spectrum.

Both electrons and muons are used in this analysis.  We combine information
 from specific ionization, energy/momentum measurements from the calorimeter
and tracking systems, and position matching from these two systems to identify
electrons down to 600 \mevc.  Muon candidates must register hits in muon
counters at least 5 interaction lengths deep, limiting the muon momentum range
to approximately 1.4 \gevc.  The probability that a hadron is misidentified as
a lepton (a ``fake lepton'') is of the order 0.1\% (1\%) for electron (muon)
identification.

Candidate $2\pi$ ($3\pi$) combinations must have an invariant mass within 90
(30) MeV of the nominal $\rho$ ($\omega$) mass.  A $\pi^0$ candidate must have
a 2-photon invariant mass within 2 standard deviations (about 12 MeV) of the
$\pi^0$ mass.  Within any one of the five modes, we pick the meson candidate in
each event that yields the smallest value of $|\dele|$.

\begin{figure}[t]
\centering
\leavevmode
\epsfbox{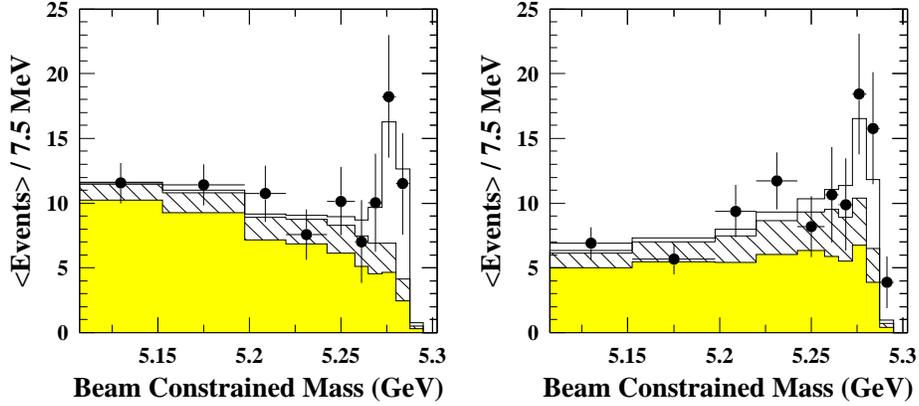}
\caption{Beam constrained mass distributions for the combined $\pimlv$ and
$\pinlv$ modes (left) and the combined vector modes (right).  The points are
continuum- and fake-subtracted data. The histograms show the contribution from
$b\to c$ (shaded), $u\ell\nu$ crossfeed (hatched) and signal (hollow).}
\label{fig:masses}
\end{figure}

We require the lepton, neutrino and meson candidates to satisfy $-250 \mbox{
MeV}<\dele<150\mbox{ MeV}$.  The cut is asymmetric because the $b\to c$
backgrounds increase rapidly as $\dele$ increases.  The range $5.265 \mbox{
GeV}<\mb<5.2875 \mbox{ GeV}$ defines the signal region.  The $\mb$ distribution
for data after all cuts, including the $\dele$ cut, is shown in
Figure~\ref{fig:masses} for the combination of the $\pimlvc$ and $\pinlvc$
modes, and for the combination of the three vector modes. There is a clear
excess above the background in the signal region for the $\pilv$ modes.  The
fit yielding the background levels shown is described in the next section.

The dominant background in both the $\pi$ and the vector modes comes from
$\btoc$ decays in events containing either an undetected $K_L$ or a second
neutrino.  The small backgrounds in each mode from fake leptons and from
continuum processes are measured with the data.  In the $\pi$ modes, Monte
Carlo studies indicate that feed-across from the $\rholv$ modes should
contribute the next largest background.  In the vector modes, $\btou$ decays to
higher mass and non-resonant final states form the other major background
component.  Our fits do not make any requirement on the distribution of events
inside our mass windows, so resonant and non-resonant final states are not
distinguished. Consequently only an upper limit on $\bbrho$ will be
obtained. We derive the limit conservatively by assuming zero background from
the non-resonant and higher mass decays.

\section{Extracting the Yields}
\begin{table}[t]
\centering
\caption{Backgrounds, efficiencies and fit results for the $\pilv$ analysis.
The $\chi^2$ for the fits using the ISGW and WSB signal models were
respectively 10.8 and 10.3 for $20-7$ degrees of freedom.  Note that the errors
on the signal yields and crossfeed backgrounds in the $\pimlv$ and $\pinlv$
modes are completely correlated because of the isospin constraints.}
\label{tab:pifit}
\begin{tabular}{lcccc} \hline\hline
                 & \multicolumn{2}{c}{$\pimlv$} &
\multicolumn{2}{c}{$\pinlv$} \\ \cline{2-5}
                 & ISGW         & WSB           &
ISGW & WSB \\  \hline
Raw Data         & \multicolumn{2}{c}{30}          &
\multicolumn{2}{c}{15} \\
Continuum Bkg.   & \multicolumn{2}{c}{$2.3\pm0.8$} &
\multicolumn{2}{c}{$1.0\pm0.5$} \\
Fake Lepton Bkg. & \multicolumn{2}{c}{$1.2\pm0.3$}   &
\multicolumn{2}{c}{$0.7\pm0.2$} \\
other $u\ell\nu$ Bkg. & \multicolumn{2}{c}{0.6} & \multicolumn{2}{c}{0.2} \\
Efficiency       & 2.9\%        & 2.1\%         & 1.9\%       & 1.4\% \\
Signal Yield     & $15.6\pm5.3$ & $16.3\pm5.3$  & $5.0\pm1.7$ & $5.3\pm1.7$ \\
$b\to c$ Bkg.    & $9.8\pm1.1$  & $9.8\pm1.1$   & $1.8\pm0.5$ & $1.7\pm0.5$ \\
$\rho/\omega$ Bkg.&$3.8\pm1.7$  & $3.4\pm1.4$   & $1.8\pm0.8$ &
$1.6\pm0.7$ \\ \hline
\end{tabular}
\end{table}

After subtracting the continuum and fake lepton backgrounds, we fit the
beam-constrained mass distributions in our five reconstructed $\btou$ modes
simultaneously, which allows the data in the vector modes to constrain the
$\rholv$ background in the $\pilv$ modes.  In addition to the signal shapes and
the feed-across shapes between the five modes, the fit includes $b\to c$ and
other $\btou$ background components. The isospin relations
$\frac12\Gamma(B^0\to\pimlv) = \Gamma(B^+\to\pinlv)$ and
$\frac12\Gamma(B^0\to\rhomlv) = \Gamma(B^+\to\rhonlv) \approx
\Gamma(B^+\to\omelv)$ constrain the neutral meson rates relative to the charged
meson rates.  We therefore obtain two yields, $\npi$ and $\nrho$, from the fit.

The $\btoc$ and feed-across background shapes in $\mb$ are obtained from Monte
Carlo simulation.  The $\btoc$ background level floats independently in each of
the five modes, while the feed-across rates between the five modes are tied to
the signal yields $\npi$ and $\nrho$.

Monte Carlo simulation also provides the $\mb$ distributions for the
non-resonant and higher mass $\btou$ backgrounds. The inclusive lepton yield at
high momentum fixes this background level.  We vary the physical model and the
rate by hand to estimate the systematic uncertainty in this procedure.

The results of the fit from which the $\pilv$ yield (and the background levels
in Figure~\ref{fig:masses}) is obtained are summarized in
Table~\ref{tab:pifit}. The efficiencies and crossfeed probabilities have been
determined using the ISGW and WSB models.  We obtain similar $\pilv$ yields for
the two models, but obtain efficiencies that differ by approximately 30\%. The
$\btoc$ background levels in the five modes are all consistent with absolute
Monte Carlo predictions based on the luminosity.  Correcting for acceptance and
averaging the electron and muon samples, we obtain the preliminary branching
fraction $\bbpi=[1.19\pm0.41]\e{-4}$ ($[1.70\pm0.55]\e{-4}$) for the ISGW (WSB)
model, where the errors are statistical only.  We obtain consistent results if
we fit using the $\dele$ distributions, having resolved multiple candidates
using $\mb$.

To obtain upper limits for the vector modes, we perform a similar fit assuming
no non-resonant or higher mass $\btou$ backgrounds.  This fit gives the
same $\pilv$ yield.  We obtain the efficiency-corrected numbers
of $834\pm337$ ($1248\pm484$) $\rhomlvc$ decays for the ISGW (WSB) model, and a
$\gamrho/\gampi$ ratio of $1.56^{+1.29}_{-0.76}$ ($1.63^{+1.21}_{-0.75}$).

\begin{figure}[b]
\centering
\leavevmode
\epsfbox{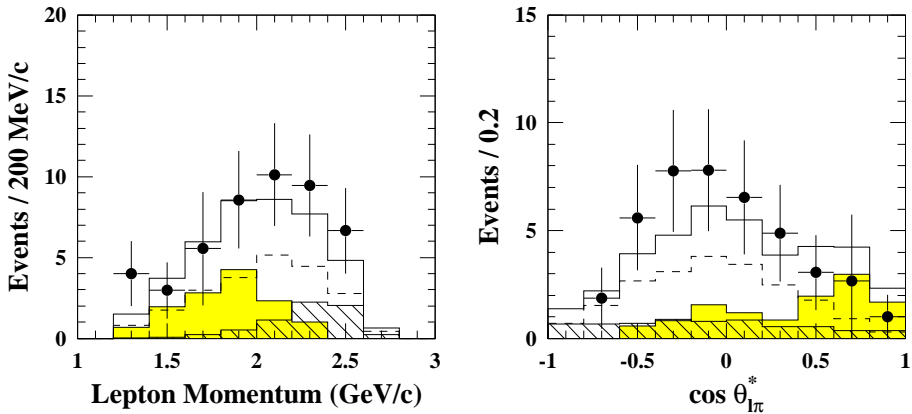}
\caption{Charged lepton spectrum (left) and $\cospil$ distribution (right) for
the combined $\pimlv$ and $\pinlv$ modes.  The
points are continuum- and fake-subtracted data. The top histogram is the total
prediction using rates from the yield fit, with components $b\to c$ (shaded),
$u\ell\nu$ crossfeed (hatched) and signal (dashed).}
\label{fig:lepspec}
\end{figure}

Many distributions have been examined for consistency with the $\pilv$
hypothesis.  The charged lepton momentum spectrum for $\pimlvc$ and $\pinlvc$
candidates in the $\mb$ signal region is shown in Figure~\ref{fig:lepspec}.
The spectrum obtained from the data is quite stiff, with a sizeable fraction of
events beyond the $\btoc$ endpoint.  The sum of the signal and background
distributions, scaled according to the fit results, shows good agreement with
the data.  The $\pi$ and $\nu$ momentum spectra are also consistent with the
results of the fit.

For $B\to\pilv$, the $V-A$ interaction predicts that the angle between the
$\pi$ and the lepton in the $W$ rest frame, $\thepil$, should have a
$\sin^2\thepil$ distribution.  The observed $\cospil$ distribution, also shown
in Figure~\ref{fig:lepspec}, is in good agreement with this expectation.  We
estimate the probability, including systematic uncertainties, that the
background processes could fluctuate to give the observed $\mb$ and $\cospil$
distributions in the combined $\pilv$ modes, and obtain $6.4\e{-5}$.  This
corresponds to a 3.8 standard deviation significance for a Gaussian
distribution.

\section{Systematics}

\begin{table}[t]
\centering
\caption{Summary of systematic uncertainties on the yields and efficiencies in
the $\pilv$ and $\rholv$ modes.  The numbers in parentheses in the background
levels indicate the uncertainty in the background as a fraction of that
background.}
\label{tab:syst}
\begin{tabular}{lrr||lrrr} \hline\hline
On yields:               & $\pilv$       & $\rholv$    &
On Efficiencies:         & $\pilv$       & $\rholv$    & $\rho/\pi$ ratio
 \\ \hline

$b\to c$ bkg.            & (20\%) 13\%   & (20\%) 23\% &
$\nu$-measurement        & 15\%          & 15\%        & 15\% \\

$\rho/\omega\ell\nu$ bkg.& (36\%)\eo\ 8\%& (63\%)\et\ 7\% &
$\pi/\rho/\omega$ finding& 3\%           & 6\%         & 7\% \\

other $u\ell\nu$ bkg.    & 8\%           & ---         &
$\rho/\omega$ polarization & ---         & 10\%        & 10\% \\

cont.+fake bkg.          & (20\%)\eo\ 6\%& (24\%)\et\ 7\% &
lepton fake rates        & 4\%           & 4\%         & 4\% \\

lepton finding           & 2\%           & 2\%         &
lepton finding           & 4\%           & 4\%         & 4\% \\

                         &               &             &
Luminosity               & 2\%           & 2\%         & --- \\ \hline

{\bf Total}              & {\bf 18\%}    & {\bf 25\%}  &
{\bf Total}              & {\bf 16\%}    & {\bf 20\%}  & {\bf 20\%} \\ \hline
\end{tabular}
\end{table}

\begin{figure}[b]
\centering
\leavevmode
\epsfbox{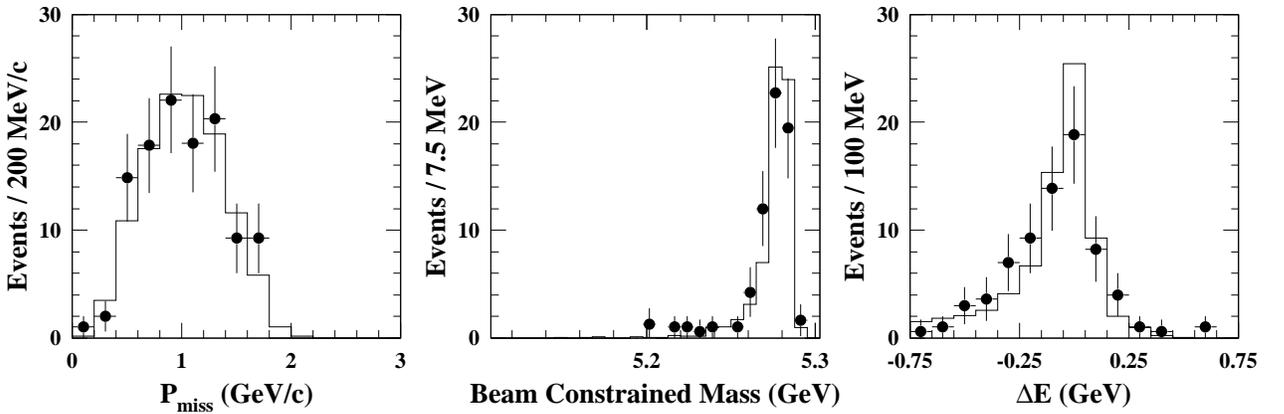}
\caption{$|\pmiss|$ spectrum (left) $\mb$ distribution (center) and $\dele$
distribution (right) for $D^{*\pm}\ell^\mp\nu$ reconstruction.  The points are
continuum and combinatoric background-subtracted data. The histograms are
signal Monte Carlo distributions normalized to equal area.}
\label{fig:dstar}
\end{figure}

The systematic uncertainties on the yields and efficiencies are summarized in
Table~\ref{tab:syst}.  The dominant uncertainty in the yields comes from the
uncertainty in the shapes of the background $\mb$ distributions.  The shapes
have been checked in a variety of ways: examining the shapes in $\dele$
sidebands and in signal-free modes (eg., $K_S\ell\nu$), and varying the
misreconstruction behavior of the Monte Carlo simulation.

The uncertainty in the efficiencies is dominated by the
neutrino-measurement simulation.  One method of estimating this uncertainty is
to use this technique to measure the branching fraction for $B\to
D^{*\pm}\ell^\mp\nu$ via the modes $D^{*\pm}\to\pi^\pm D^0$, $D^0\to
K^\mp\pi^\pm$. We find that the simulation of $\emiss$, $\pmiss$, $\dele$ and
$\mb$ agrees well with the data (Figure~\ref{fig:dstar}), and that ${\cal
B}(B^0\to D^{*-}\ell^+\nu)=4.66\pm0.65\%$. This agrees with the published CLEO
result \cite{bb:dstar} of $4.49\pm0.32\pm0.32\%$, which used a
higher statistics technique.  The 15\% statistical uncertainty is taken as the
systematic uncertainty; other studies indicate that this is a conservative
estimate.  We expect this systematic to cancel in the $\rho/\pi$ ratio,
but retain a preliminary 15\% uncertainty.

\section{Conclusion}
Combining the $\pilv$ yields and the systematic uncertainties, we obtain the
preliminary branching fraction ${\cal
B}(B^0\to\pimlv)=[1.19\pm0.41\pm0.21\pm0.19]\e{-4}$
($[1.70\pm0.51\pm0.31\pm0.27]\e{-4}$) using the ISGW (WSB) model to evaluate
efficiencies.  The errors are statistical, systematic on the yield, and
systematic on the efficiency, respectively.  This is the first measurement of
any exclusive $\btou$ branching fraction.  The probability
of a background fluctuation resulting in the observed signal is $6.4\e{-5}$.

Assuming no non-resonant or high mass $u\ell\nu$ background, we obtain a
conservative 90\% C.L. upper limit of ${\cal B}(B^0\to\rhomlv) < 3.1\e{-4}$ for
the ISGW model and ${\cal B}(B^0\to\rhomlv) < 4.6\e{-4}$ for the WSB model.
The statistical and systematic uncertainties have been combined in quadrature
in evaluating these limits.  The results are comparable to the previous CLEO
upper limits for the vector modes.

Finally, we find $\gamrho/\gampi<3.4$ at the 90\% confidence level for both
the ISGW and WSB models.  Again, statistical and systematic uncertainties have
been combined in quadrature.  Comparing to the predictions in
Table~\ref{tab:exclpred}, the WSB model is compatible with this limit, but it
is difficult to reconcile the ISGW model with this limit.

These preliminary measurements herald a new era for the study of $\vub$.  CLEO
is still refining these measurements, with 50\% more data soon to be available
and work in progress on the separation of the vector modes from non-resonant
modes.

\section{References}

\end{document}